\documentclass[12pt,preprint]{aastex}
\usepackage{emulateapj5}

\shorttitle{X-RAY BINARIES AND GLOBULAR CLUSTERS IN ELLIPTICAL GALAXIES}
\shortauthors{WHITE, SARAZIN, \& KULKARNI}

\slugcomment{The Astrophysical Journal Letters, submitted}

\begin{document}

\title{
X-ray Binaries and Globular Clusters in Elliptical Galaxies
}

\author{
Raymond E.~White III\altaffilmark{1},
Craig L.~Sarazin\altaffilmark{2},
and Shrinivas R.~Kulkarni\altaffilmark{3}
}
\altaffiltext{1}{
Department of Physics and Astronomy, University of Alabama,
Box 870324, Tuscaloosa, AL 35487-0324; rwhite@bama.ua.edu
}
\altaffiltext{2}{
Department of Astronomy, University of Virginia,
P.~O.~Box 3818, Charlottesville, VA 22903-0818; cls7i@virginia.edu
}
\altaffiltext{3}{
California Institute of Technology, Owens Valley Radio Observatory,
105-24, Pasadena, CA 91125; srk@astro.caltech.edu
}
\begin{abstract}
The X-ray emission from normal elliptical galaxies has two major components:
soft ($kT\approx0.2-1$ keV) emission from diffuse gas and harder
($kT\approx6$ keV) emission from populations of accreting (low-mass)
stellar X-ray binaries (LMXB).
If the LMXB population is intimately tied to the field stellar
population in a galaxy, its aggregate X-ray luminosity is expected to be
simply proportional to the optical luminosity of the galaxy.
However, recent {\sl ASCA} and {\sl Chandra} X-ray observations
show that the global luminosities of LMXB components in ellipticals exhibit
significant scatter (a factor of $\sim$4) at a given optical luminosity.
This scatter may reflect a range of evolutionary stages among X-ray binary
populations in elliptical galaxies of different ages.
If so, the ratio of the global LMXB X-ray luminosity to the galactic optical
luminosity, $L_{\rm LMXB}/L_{\rm opt}$, may in principle be used to
determine when the bulk of stars were formed in individual ellipticals.
To test this, we compare variations in $L_{\rm LMXB}/L_{\rm opt}$ for
LMXB populations in ellipticals to optically-derived estimates of stellar
ages in the same galaxies.
We find no correlation, which suggests that variations in
$L_{\rm LMXB}/L_{\rm opt}$ are not a good age indicator for ellipticals.
Alternatively, LMXBs may be formed primarily in globular clusters
(through stellar tidal interactions), rather than comprise a subset of
the primordial binary star population in a galactic stellar field.
Since elliptical galaxies exhibit a wide range of globular cluster populations
for a given galaxian luminosity, this may induce a dispersion in the
LMXB populations of ellipticals with similar optical luminosities.
Indeed, we find that $L_{\rm LMXB}/L_{\rm opt}$ ratios for LMXB populations 
are strongly correlated with the specific globular cluster frequencies in 
elliptical galaxies.
This suggests that most LMXBs were formed in globular clusters.
If so, {\sl Chandra} observations of central dominant
galaxies with unusually large globular cluster populations should
find proportionally excessive numbers of LMXBs.
\end{abstract}


\section{Introduction} \label{sec:intro}

The X-ray emission from normal elliptical galaxies has two major components:
soft ($kT\approx0.2-1$ keV) emission from diffuse gas and harder
($kT\approx6$ keV) emission from
populations of accreting (low-mass) stellar X-ray binaries (LMXB).
The X-ray properties of the LMXB component have been difficult to determine,
due to their spatial confusion with diffuse gaseous emission and to
spectral hardness which places much of the LMXB emission outside
the effective bandpasses of most imaging X-ray satellite spectrometers.
The presence of the LMXB component has been inferred in part through
observations of spectral hardening in ellipticals with progressively smaller
X-ray to optical luminosity ratios (Kim, Fabbiano, \& Trinchieri 1992),
indicating that they have relatively little gas, exposing the harder
LMXB component.
Populations of LMXBs are also expected in ellipticals simply by
analogy with detections of discrete LMXB sources in nearby spheroids
such as the bulges and halos of our Galaxy and M31
(Forman, Jones \& Tucker 1985; Canizares, Fabbiano \& Trinchieri 1987),
as well as in the radio galaxy Centaurus A (Turner et al.\ 1997).

A simple argument suggests that the total X-ray luminosities of
LMXB populations in ellipticals might be proportional to the 
stellar luminosities of the galaxies:
if the properties of low mass binary stellar systems
(such as the fraction of stars in binaries, the distributions of binary
separations and mass ratios, etc.) are largely independent of their galactic
context, the number of LMXBs (hence their aggregate X-ray luminosity)
should be simply proportional to the number of stars in the galaxy
(and thus their total optical luminosity).

High angular resolution {\sl Chandra} observations are now allowing individual
LMXBs to be resolved out of the diffuse gaseous X-ray emission in nearby
ellipticals
(Kraft et al.~2000, 2001;
Sarazin, Irwin, \& Bregman 2000, 2001;
Angelini, Loewenstein, \& Mushotzky 2001;
Finoguenov \& Jones 2001),
which makes their composite spectral analysis much easier.
The bulk of the hard emission in normal ellipticals indeed comes from LMXBs,
rather than from advection-dominated accretion flows onto massive,
central black holes, proposed by Allen, di Matteo \& Fabian (2000).

Until {\sl Chandra} observes more nearby ellipticals,
the strongest spectral constraints to date on the hard stellar LMXB component
in a large sample of ellipticals will still come from {\sl ASCA}
spectra  (Matsumoto et al.~1997; White 2000, 2001, 2002).
Since the effective angular resolution of {\sl ASCA} imaging spectrometers is
$2-3'$ (half-power diameter), confusion prevents individual LMXBs from
being easily resolved out of the diffuse gas in ellipticals.
The hard LMXB component can be spectrally distinguished from the softer
gaseous component, however.
Matsumoto et al.~(1997) separated the hard LMXB component in ellipticals
from softer gaseous emission by considering {\sl ASCA} GIS spectral energies
above 4 keV.
Stacking 4-10 keV GIS spectra for 12 ellipticals,
they found a best-fit thermal bremsstrahlung model
with $kT=12_{-5.5}^{+29.3}$ keV (where errors are 90\% confidence limits);
a power-law with photon index $\alpha=1.8_{-0.4}^{+0.4}$ fit equally well.
Matsumoto et al.~(1997) found that the X-ray luminosities of the LMXB components
were proportional to the optical luminosities of the ellipticals, but with a
surprisingly large scatter (a factor of $\sim4$).
Some of the ellipticals included in the sample have significant X-ray emission
from active galactic nuclei (AGN), which may account for some of the scatter.

White (2000, 2001, 2002) performed a similar {\sl ASCA} analysis, but with a larger
spectral bandwidth (0.7-10 keV), on six normal ellipticals
(i.e., ellipticals without significant AGN emission).
Spectra were extracted from within metric radii of 6$r_e$ from the galactic
centers, where $r_e$ is the optical effective radius of a galaxy.
The GIS spectra of the six ellipticals were simultaneously fit with both
soft (gaseous) and hard (LMXB) emission components.
(Only GIS data were used because the GIS detectors have twice the 
effective area of the SIS detectors above 7 keV.)
The temperatures (or power-law indices) of the hard components in the
galaxies were tied together, while the temperatures of the soft components
(if present) were allowed to vary individually.
Much tighter spectral model constraints were provided by the increased
spectral bandwidth compared to the 4-10 keV bandwidth in the stacked spectral
study of Matsumoto et al.~(1997).
The spectra of the LMXB components were fit equally well by a
bremsstrahlung model with $kT = 6.4_{-1.1}^{+1.7}$ keV or a
power-law model with photon index $\alpha = 1.82_{-0.09}^{+0.10}$
(errors are 90\% confidence limits).
Individual fits to each galaxy in the set were consistent with the 
results of the joint fits and fluxes were obtained by adopting the
best jointly fit temperature.
These are the tightest constraints to date on the global spectral properties
of the stellar LMXB component in ellipticals.
GIS X-ray fluxes were determined for the LMXB components in an additional 
8 ellipticals which had poorer photon statistics by fixing the 
temperature of the hard component to 6.4 keV.
The resulting X-ray luminosities of the LMXB components in the 14 galaxies
were found to be proportional to the optical luminosities of the galaxies,
with a factor of $\sim4-10$ scatter.

Recent {\sl Chandra} observations
(Sarazin et al.~2000, 2001:
Angelini et al.\ 2001;
Blanton, Sarazin, \& Irwin 2001)
also show that the X-ray luminosities of the resolved LMXB components in
ellipticals exhibit significant scatter at a given optical luminosity.
(The composite spectra of these LMXB components are also consistent with 
the hard {\sl ASCA} spectral results described above.)
Although their scatter in luminosity is much smaller than that of the 
softer gaseous component (which ranges a factor of 100 in X-ray/optical flux
ratio), it is still larger than expected if the LMXB component is
strictly proportional to the stellar component.
Following a preliminary study by White (2000, 2001),
we next consider two possible sources of variance in the X-ray luminosities of
LMXB populations in elliptical galaxies: age differences among ellipticals and
differences in their globular cluster populations.

\section{Possible Sources of Variance in LMXB Populations in Ellipticals}
\label{sec:tests}

\subsection{Variations in Galaxy Ages} \label{sec:tests_age}

If LMXB populations are a subset of the primordial binary systems
in a galactic stellar field, then the evolution of an LMXB population
should be tied to the evolution of visible stars.
The evolution of the aggregate X-ray luminosity of an LMXB population
is best assessed with a population synthesis calculation (Wu 2001).
A simpler, more phenomenological approach was adopted by
White \& Ghosh (1998).
The luminosity of an LMXB population (neglecting burst sources) is
expected to rise for about a Gyr (White \& Ghosh 1998),
driven by the time it takes the less massive secondaries in binaries to
evolve off the main sequence, overflow their Roche lobes, and start dumping
mass onto the more massive (primary) compact stellar remnants in the binaries.
After this Gyr ramp-up, the luminosity of an LMXB population is expected
to slowly decline (White \& Ghosh 1998; Wu 2001).

\centerline{\null}
\vskip2.45truein
\includegraphics{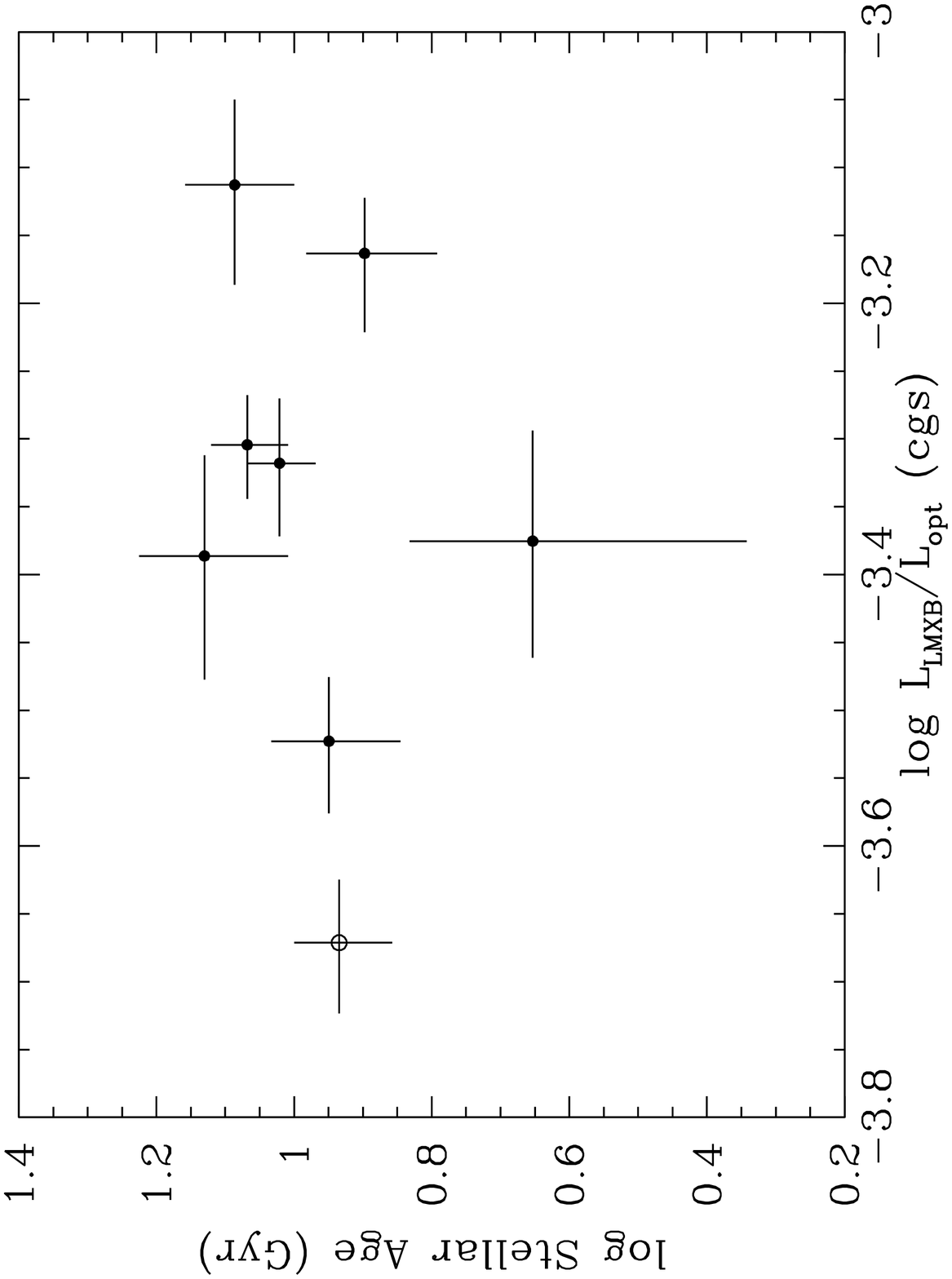}
\figcaption{Mean ages of the stars in elliptical galaxies
(Trager et al.\ 2000)
are plotted against the ratios of the X-ray luminosity of their LMXBs to their
optical $B$-band luminosity (White 2002). The open circle indicates an upper
limit to $L_{\rm LMXB}$ due to the presence of an AGN.
\label{fig:age}}

The optical stellar luminosities of galaxies are also slowly declining with time.
Figure~1 of Bruzual \& Charlot (1993) shows that the (visual)
optical luminosity of an elliptical galaxy declines as
$L_V\propto t^{-0.95}$.
This time dependence is close enough to estimates for the temporal
decline in the X-ray luminosity of LMXB populations that it is difficult
to assess how the X-ray to optical ratio of the LMXB component is evolving
at the present epoch.

For our purposes,  ambiguity in
the theoretical expectation for the temporal evolution of
$L_{\rm LMXB}/L_{\rm opt}$
does not matter --- we can deduce the evolution empirically.
If variations in the X-ray properties of the LMXB component
for galaxies of a given optical luminosity are correlated with
independent age estimates, the $L_{\rm LMXB}/L_{\rm opt}$ ratio itself
might then be used as an indicator of galactic age.

To test this possibility, we compare the $L_{\rm LMXB}/L_{\rm opt}$ ratio for
the LMXB components from the elliptical sample of White (2002) to
recent stellar age estimates for elliptical galaxies (Trager et al.~2000).
The stellar age estimates were derived by comparing the strengths of optical
stellar spectral line indices of $<$Fe$>$, Mg~b and H$\beta$ (which
were corrected for the effects of non-solar abundance ratios), to
the single-burst stellar population models of Worthey (1994).
For their sample of 50 ellipticals, Trager et al.~(2000) find ages ranging
from $1.5-18$ Gyr.
In Figure~\ref{fig:age},
the relevant optical stellar age estimates (ranging from
$\sim4-14$ Gyr) are compared to the $L_{\rm LMXB}/L_{\rm opt}$ ratios for
the LMXB populations in the eight ellipticals of (the fourteen in)
White (2002) that were included in the Trager et al.~(2000) sample.
The X-ray luminosities were calculated in the 0.5-4.5 keV range and their
errors are 90\% confidence limits.
There is apparently no correlation, so variations in the
$L_{\rm LMXB}/L_{\rm opt}$ ratios for LMXB populations are not likely to be
due to galactic age differences.

\vskip0.2truein

\subsection{Globular Cluster Population Variations} \label{sec:tests_gc}

Alternatively, most LMXBs may be formed in globular clusters, rather
than in the stellar field of a galaxy.
Observations of LMXBs in our own Galaxy show that it is very difficult
to make LMXBs in the field.
There are comparable numbers of high mass X-ray binaries (HMXBs)
and LMXBs, but HMXB lifetimes are only $\sim10^5$ yr, which are
$10^{2-5}$ times shorter than LMXB lifetimes.
Thus, the formation of LMXBs in our Galaxy is $10^{2-5}$ times less
efficient than the formation of HMXBs.

Meanwhile, globular clusters are the most efficient formation
site for LMXBs in our Galaxy:
$\sim20\%$ of all LMXBs in our Galaxy reside in globular clusters,
yet globular clusters contain $\lesssim0.1\%$ of the stars in our Galaxy
(Katz 1975).
Thus, globular clusters form LMXBs $>200$ times more efficiently
than the rest of the Galaxy, through tidal capture
during close stellar encounters (Clark 1975).

Given this, it is reasonable to consider whether
$most$ LMXBs are formed in globular clusters.
Grindlay (1984) suggested that X-ray bursting binaries in the bulge
of our Galaxy (not residing in globular clusters)
were formed in globular clusters which were later destroyed by tidal
shocks while passing through the Galaxy's disk.
This disruption mechanism is not relevant to
globulars residing in elliptical galaxies, but if globulars are on eccentric
orbits, they may still lose stars due to galactic tides at perigalacticon.
Vesperini (2000) estimates that only $\sim10\%$ of globular clusters
are disrupted in the most luminous ellipticals, but the fraction disrupted 
increases considerably for progressively smaller elliptical luminosities.
Also, some LMXBs formed in globular clusters 
may be ejected by the velocity kicks imparted by supernovae during
the formation of neutron stars.
Typical supernovae kick velocities are thought to range from $100-500$ km/s
(Kalogera 1996; Terman, Taam, \& Savage 1996; Fryer, Burrows, \& Benz 1998),
while the escape velocities of globular clusters range
from $5-60$ km/s (Webbink 1985).
Since LMXB lifetimes can be quite long ($\gtrsim10^9$ yr), LMXBs ejected from
globular clusters can be visible outside the clusters for a long time.
If LMXBs are made primarily in globular clusters, we would expect
their number to be proportional to the number of globular clusters,
regardless of the fraction ejected from clusters.

With the excellent angular resolution of {\sl Chandra},
it is possible to determine positions of LMXBs with sufficient
accuracy to determine if they coincide with globular clusters.
Observations of ellipticals show that a significant fraction
are located in globulars
(Sarazin et al.\ 2000, 2001;
Angelini et al.\ 2001;
Randall, Sarazin, \& Irwin 2001;
White, Davis \& Hanes 2002).
The fraction ranges from $\ga$20 to 70\%.
These are lower limits, since deeper {\sl Chandra} exposures may
detect fainter X-ray sources in additional globulars.
In the same region of these galaxies, the globular clusters typically
provide only $\sim 0.1$\% of the optical light, which implies that
optical stars in globular clusters are much more likely to be the donor
stars in LMXBs than in the field.
The primary limitation with identifying LMXBs with globulars in ellipticals
at present is the lack of {\it Hubble Space Telescope} (HST) globular cluster
lists for many of the {\it Chandra} observed ellipticals.

Since elliptical galaxies exhibit an order of magnitude range of globular
cluster populations for a given galaxian luminosity,
we can assess whether these variations are
correlated with variations in $L_{\rm LMXB}/L_{\rm opt}$.
In Figure~\ref{fig:gc},
we compare the specific frequency of globular clusters $S$
(the number of globular clusters per galaxy visual luminosity in units of
the luminosity of a $M_V = -15$ galaxy) from the compilation of
Kissler-Patig (1997) to
the $L_{\rm LMXB}/L_{\rm opt}$ ratio for the LMXB populations in the elliptical
sample described above.
The Kissler-Patig (1997) compilation includes error estimates for
$S$, as indicated in the figure, for all but one elliptical in this sample.
There appears to be a strong correlation (White 2000, 2001), 
and the relationship is consistent
with direct proportionality: $L_{\rm LMXB}/L_{\rm opt}\propto S^{1.2\pm0.4}$.
Thus, LMXB populations may indeed be controlled by globular cluster populations.

\centerline{\null}
\vskip2.45truein
\includegraphics{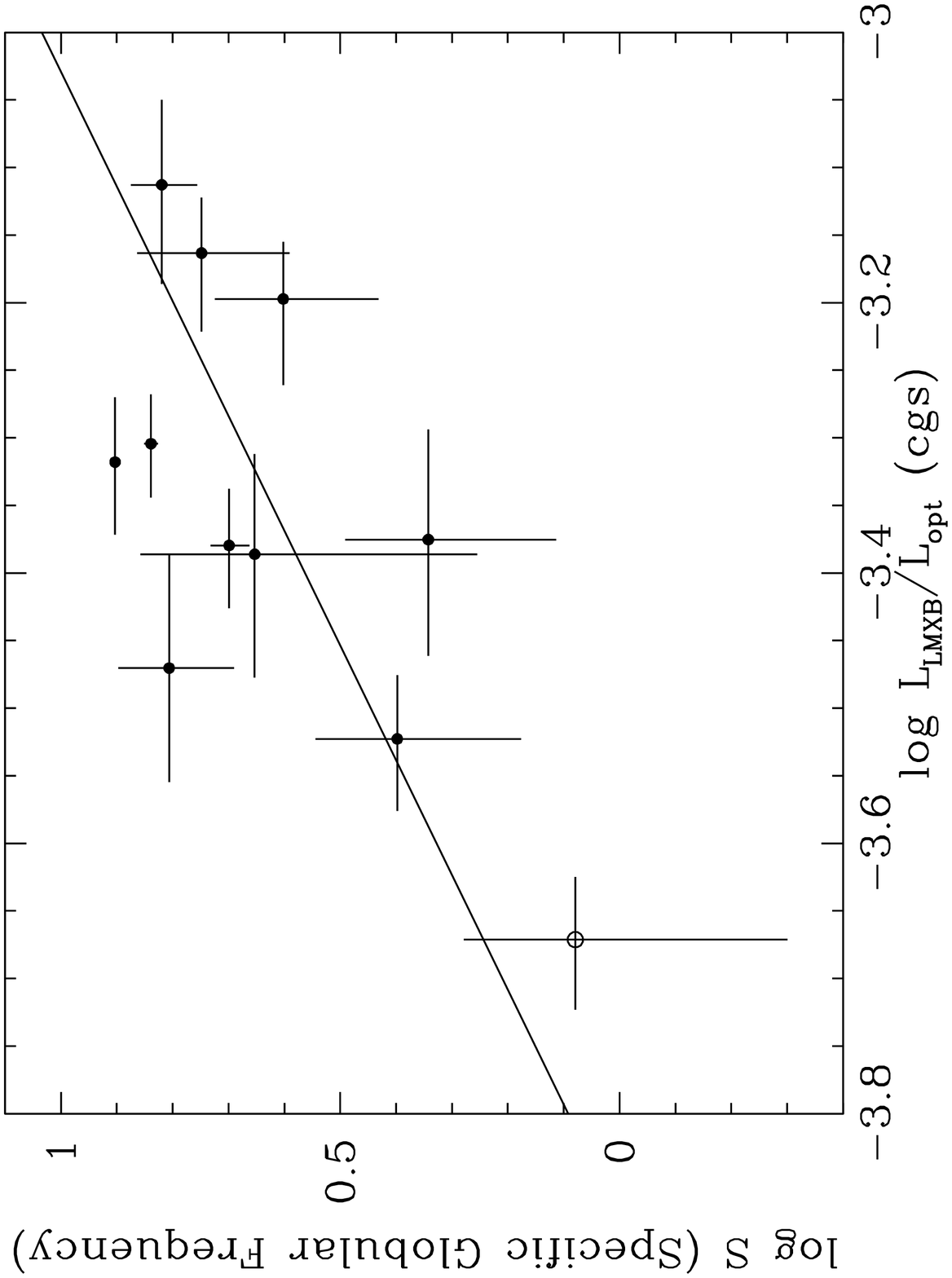}
\figcaption{Logarithm of the specific frequency of globular clusters $S$
(Kissler-Patig 1997) vs.\ the X-ray/optical luminosity ratio for elliptical
galaxies.
The best fit relationship (using the ASURV statistical package
(Isobe et al 1986) is
$L_{\rm LMXB} / L_{\rm opt} \propto S^{1.2 \pm 0.4}$, with a 90\% probability
of correlation via Kendall's $\tau$ test.
The open circle indicates an upper
limit to $L_{\rm LMXB}$ due to the presence of an AGN.
\label{fig:gc}}

\vskip0.2truein

\section{Discussion and Conclusions} \label{sec:conclusion}

This study suggests that globular cluster populations in ellipticals
control the evolution of their populations of low mass stellar X-ray binaries.
At the present epoch, age differences among the ellipticals
in the small sample described above do not seem to be correlated with the
variations in their X-ray binary populations.
It would be useful to compare the LMXBs in early-type galaxies with
more detailed properties (such as color and magnitude) of their globulars
(other than just the number),
to see if the history of star formation in the globular affects the LMXB population.
The {\it Chandra} observations of NGC~1399 suggested that LMXBs are more
likely to be found in redder globular clusters (Angelini et al.\ 2001).
There also is evidence that LMXBs are more likely to be found in brighter globulars
(Angelini et al.\ 2001),
although it is unclear whether this is just due to the larger
number of stars in brighter globulars, or indicates a higher probability
per star in more massive clusters.

It is possible that most or all LMXBs in early-type galaxies are formed
in globular clusters.
In this scenario, any field LMXBs in ellipticals would have escaped from globulars.
The field LMXBs might have been ejected by the kick velocities resulting
from supernovae, by stellar dynamical processes such as superelastic
encounters in which internal binding energy in the binaries is converted
into kinetic energy of motion of the binary, or by the dissolution of the
globular due to tidal effects.

There is some evidence that the LMXBs in globular clusters in elliptical
galaxies are brighter in X-rays than those in the field
(Angelini et al.\ 2001).
If all of the LMXBs were formed in globular clusters, this would require
that the less luminous and presumably less massive LMXBs systems were
preferentially ejected from globular clusters.
Alternatively, if a significant fraction of the field LMXBs were formed
in the field, this would imply differing stellar and binary evolution
histories in the field and in globular clusters.

Many sources associated with globulars have X-ray luminosities which
exceed the Eddington luminosity of a 1.4 $M_\odot$ neutron star,
$L_{\rm Edd,NS} \approx 2 \times 10^{38}$ ergs s$^{-1}$
(Sarazin et al.\ 2000, 2001;
Angelini et al.\ 2001;
Blanton et al.\ 2001;
Randall et al.\ 2001;
White et al.\ 2002).
In general, LMXB luminosity functions within early-type galaxies have
a break at a luminosity very close to $L_{\rm Edd,NS}$
(Sarazin et al.\ 2000, 2001;
Blanton et al.\ 2001;
Randall et al.\ 2001).
This break may separate accreting
black holes, at higher luminosities, from accreting neutron stars
at lower luminosities (Sarazin et al.\ 2000).
If the sources are Eddington-limited single binaries, then
the brightest sources associated with globular clusters must contain quite
massive ($\ga 20 M_\odot$) black holes
(Sarazin et al.\ 2000).
It is not clear how globular clusters would form and
retain such massive black holes in LMXBs and also not have them form
binary black hole systems
(e.g., Portegies Zwart \& McMillan 2000).

On the other hand, these super-Eddington sources might be globular clusters
containing multiple X-ray sources
(Angelini et al.\ 2001).
The difficulty with this suggestion is that the fraction of globulars
which contain at least one X-ray source is low;
the fraction is about 3-4\% in NGC~1399 (Angelini et al.\ 2001),
NGC~4697 (Sarazin et al.\ 2000, 2001), and in NGC~4649.
(It is interesting that this fraction is almost the same in these three ellipticals.)
The luminosity function of the LMXBs isn't steep enough that multiple sources 
should affect the luminosity function at the high end, given this low fraction.
On the other hand, if the LMXBs are mainly confined to a smaller subset of
globular clusters (e.g., very bright and/or very red and/or core--collapsed
systems), then the fraction of the relevant subset of globulars with
X-ray sources and the probability of multiple sources might be much higher.

There is some evidence that these luminous super-Eddington X-ray sources are
associated with globular clusters in elliptical and S0 galaxies, but not
with the bulges of spiral galaxies.
There are no super-Eddington LMXBs associated with globular clusters in
the bulge of our Galaxy
(e.g., Hut et al.\ 1992).
The same is true of the central 5\arcmin\ of the bulge of M31
(Shirey et al.\ 2001).
The luminosity of LMXBs in the nearby large spiral bulge NGC~1291 has
a cut-off at a luminosity which is very close to $L_{\rm Edd,NS}$.
Unless this is the result of statistical fluctuations and the large
populations of globular clusters in elliptical galaxies, this suggests
that the binaries in globulars in ellipticals have different histories
than those in spiral bulges.

In general, the specific frequency of globular clusters $S$ increases along
the Hubble sequence from late to early-type spirals, from spirals to S0s,
from S0s to ellipticals, and from normal giant ellipticals to cDs
(e.g., Harris 1991).
If LMXBs all are born in globular clusters, then one would expect
the specific frequency of LMXBs and their X-ray to optical ratio to
increase along the same sequence.
In particular,
a number of central dominant ellipticals in galaxy clusters are notable
for having enormous populations of globular clusters.
White (1987) suggested that these unusually populous systems belong
to the galaxy clusters, rather than individual galaxies:
globular clusters will have been
tidally stripped from individual galaxies during the collapse of a
galaxy cluster and will then virialize along with other cluster constituents.
If LMXBs are formed primarily in globular clusters, then central dominant
galaxies with high specific globular cluster frequencies
(such as M87, NGC~1399, NGC~3311 and NGC~4874)
should be particularly rich in LMXBs, as well.
{\sl Chandra} observations have already confirmed the large populations
of LMXBs in NGC~1399
(Angelini et al.\ 2001),
and in the region of this galaxy with a globular cluster list based
on HST observations,
70\% of the LMXBs are located in globular clusters.

\acknowledgements
R.\ E.\ W.\ 
was supported in part by a National Research Council Senior Research
Associateship at NASA/Goddard and by the National Aeronautics and Space
Administration through $Chandra$ Award Numbers
GO 0-1144X and GO 1-2095X
issued by the {\it Chandra} X-ray Observatory Center, which is operated by the
Smithsonian Astrophysical Observatory for and on behalf of NASA under
contract
NAS8-39073.
C.\ L.\ S.\ was supported in part by $Chandra$ Awards
GO0-1141X and
GO1-2078X
and by {\it XMM/Newton} award
NAG5-10074.



\end{document}